\documentclass{elsart}

\usepackage{graphicx}

\usepackage{amssymb}

\begin{document}

\begin{frontmatter}



\title{An algorithm for simulating the Ising model on a type-II quantum computer}


\author{J.~H.~Cole\corauthref{cor1}},
\ead{j.cole@physics.unimelb.edu.au}
\author{L.~C.~L.~Hollenberg} \and
\author{S.~Prawer}

\corauth[cor1]{Corresponding Author}

\address{Centre for Quantum Computer Technology}
\address{School of Physics, University of Melbourne, VIC 3010,
Australia}

\begin{abstract}
Presented here is an algorithm for a type-II quantum computer which simulates the Ising model in one and two dimensions.  It is equivalent to the Metropolis Monte-Carlo method and takes advantage of quantum superposition for random number generation.  This algorithm does not require the ensemble of states to be measured at the end of each iteration, as is required for other type-II algorithms. Only the binary result is measured at each node which means this algorithm could be implemented using a range of different quantum computing architectures.  The Ising model provides an example of how cellular automata rules can be formulated to be run on a type-II quantum computer.
\end{abstract}

\end{frontmatter}

\section{Introduction}\label{intro}
\begin{figure} [htb]
\centering{\includegraphics[width=6cm]{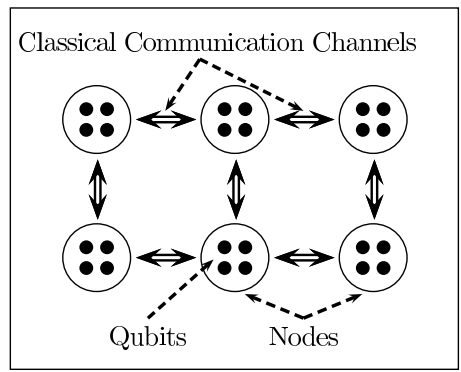}} \caption
{Design of a Type-II Quantum Computer, composed of many small conventional quantum computers\label{fig:T2QC}}
\end{figure}

There has been much effort recently in the development of quantum computing schemes for a range of applications.  One such scheme consists of an array of small quantum information processors connected by classical communication channels and is termed a type-II quantum computer (T2QC)~\cite{Yepez:01a}. Figure \ref{fig:T2QC} shows conceptually the design of a T2QC where each node contains a small number of qubits (10's or less) and the nodes are connected via classical communication channels.  Each node is initialised into some quantum state and then the entire computer undergoes some global unitary operation.  The state of each node is then measured and the results used to re-initialise the neighbouring nodes for the next computational step.  These are known as the collision and streaming steps respectively, using the lattice-gas terminology.

This form of quantum computer is able to solve certain fluid dynamics problems which are difficult to solve on a classical computer~\cite{Yepez:01b}\cite{Yepez:01c}\cite{Vahala:03}. While this is a useful application in its own right, it also provides an important `stepping stone' towards globally phase coherent quantum computers of the type required to run Shor's factoring algorithm~\cite{Shor:97} and other important quantum computing algorithms.

The type-II architecture is also well suited to a range of other physical problems. One such system is the Metropolis Monte-Carlo (MC) method for simulating the Ising model. It has been previously suggested that this problem can be simulated using a set of cellular automaton rules on a lattice~\cite{Vichniac:84}. Using this approach, the problem is well suited to a type-II quantum computer as it requires many lattice points but only nearest neighbour interactions.  While there are several other deterministic or semi-deterministic schemes available to simulate the Ising model with simple cellular automata~\cite{Creutz:86}\cite{Ottavi:89}\cite{Herrmann:86}, they are only approximations to the metropolis solution.  This is because the random number generation required by the metropolis algorithm is difficult to implement without float-point numbers or at least very large integers.  A cellular automaton rule typically requires only small numbers of bits or integers.  The major advantage of using a T2QC to implement the metropolis algorithm is that the random number generation can be included using quantum superposition. 

In this paper, the required evolution rules are developed for the Metropolis algorithm and then the necessary unitary evolution is determined using the quantum circuit formalism.  The probabilistic aspect of the this algorithm is incorporated in the form of the weighted coin-toss application of quantum superposition.  The necessary quantum evolution circuits are derived for both the one-dimensional (1D) and two-dimensional (2D) Ising models.  The implementation of these circuits is discussed for technologies where only the resulting binary state can be measured and also where the ensemble of states can be measured.

\section{Metropolis simulation of the Ising model}\label{metro}
The Ising model consists of a regular array of spins $\mathbf{s}_{i}$ which take one of two values $\{-1,1\}$. The energy at each site $i$ is given by

\begin{equation} \label{equ:IsingE}
E_{i}=-J\sum_{j} \mathbf{s}_{i}\mathbf{s}_{j}
\end{equation} 

where the sum is over nearest neighbours ($j=i\pm1$) and the total energy of the lattice is given by the sum of the on-site energies.  The use of periodic boundary conditions is assumed, though fixed boundaries can be easily implemented.  As the spins can take only two values (assuming there is no external field) the energy at each site can only take a finite number of values, the number of which depend on the dimension of the lattice.  For the case of a 1D chain of Ising spins, the change in energy can take three distinct values, $\Delta E=0J$, $\pm4J$.  The value of $J$ is assumed to be constant across the lattice and for most of the discussion, it is assumed to be positive ($J>0$).  This situation corresponds to the ferromagnetic case, though the generalisation to the anti-ferromagnetic case ($J<0$) is relatively simple.

Metropolis Monte-Carlo\cite{Metropolis:53} is a popular method of simulating the Ising model for a given temperature. The metropolis algorithm involves flipping a spin at random and calculating the corresponding change in energy ($\Delta E$).  This spin flip is then accepted with probability given by $min\{1,\e^{-\frac{\Delta E}{kT}}\}$, where $T$ is the temperature of the lattice and $k$ is the Boltzmann constant.  Throughout this discussion it is assumed that $k=1$ which results in temperature being expressed in units of $J$.

As the energy levels given by equation \ref{equ:IsingE} are discrete, the metropolis algorithm is easily converted to a set of cellular automaton rules.  In order to implement this, the same evolution rule must be applied to the entire lattice simultaneously.  However, if the metropolis algorithm is applied to every site simultaneously the `feedback catastrophe' results, as pointed out by Vichniac~\cite{Vichniac:84}.  Instead, the rule must be applied to (at most) every second site in a checkerboard configuration~\cite{Creutz:79a}. This can be achieved by either storing two sublattices or using a parity bit which controls whether the evolution rule is applied or not~\cite{Creutz:86}.

\section{Quantum circuit formalism}\label{qcct}
As the quantum circuit formalism is universal and encompasses classical boolean logic~\cite{Barenco:95}, type-II quantum computers can be designed to simulate most deterministic and probabilistic cellular automata.  The probabilistic aspect can be incorporated through the use of a qubit in a superposition state to provide a weighted coin-toss probability.  The Ising model algorithm is presented as an example of this idea, though the process should generalise for most cellular automata rules. 

The appropriate evolution rules required to implement the Metropolis algorithm are given here using the conventional quantum circuit notation~\cite{Barenco:95}\cite{Nielsen:02}. This formalism provides a simple way to describe the evolution while offering a straight forward method of checking the operation via a truth-table approach.  The notation used to represent the reversible gates used in these circuits is briefly described here.  For a more complete discussion, see chapters 1 and 4 of Nielsen and Chuang\cite{Nielsen:02}.  

\begin{figure} [htb]
\centering{\includegraphics[width=8cm]{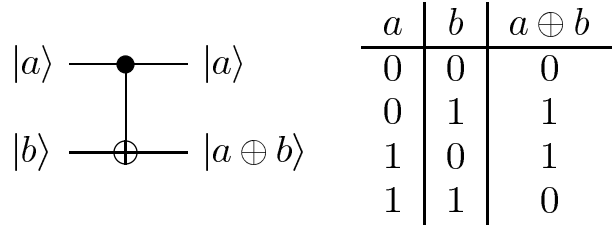}} \caption
{Notation for a CNOT gate and the truth-table showing classical operation\label{fig:CNOT}}
\end{figure}

\begin{figure} [htb]
\centering{\includegraphics[width=8cm]{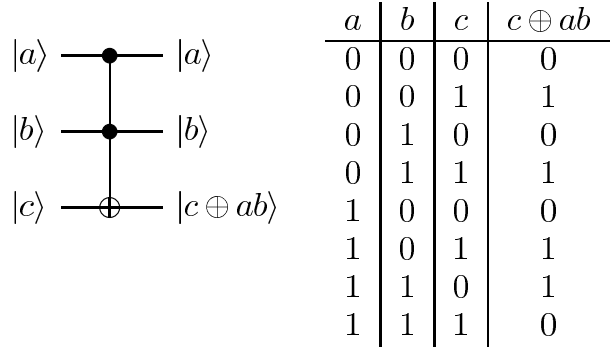}} \caption
{Notation for a Toffoli gate and the truth-table showing classical operation\label{fig:toffoli}}
\end{figure}

\begin{figure} [htb]
\centering{\includegraphics[width=8.5cm]{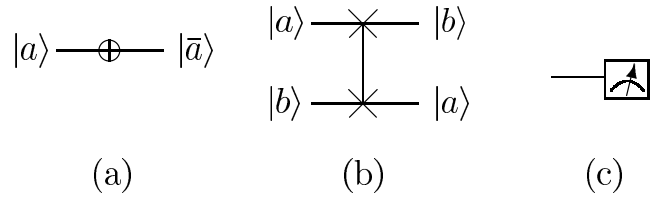}} \caption
{Notation for a NOT gate (a), a swap gate (b) and a destructive measurement (c)\label{fig:invertswap}}
\end{figure}

\begin{figure} [htb]
\centering{\includegraphics[width=9cm]{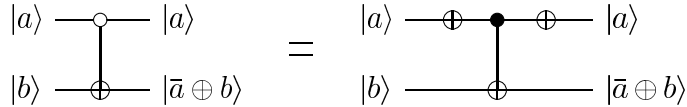}} \caption
{Notation for a CNOT gate where the target bit is inverted if the control bit is equal to $0$\label{fig:invertedcnot}}
\end{figure}

The CNOT or controlled-NOT gate is illustrated in figure \ref{fig:CNOT} and is accompanied by the corresponding truth-table.  The basic operation of the CNOT gate is to invert the target bit ($|b\rangle$) if the control bit ($|a\rangle$) is equal to one.  The convention is to use a filled circle to represent the control bit while inversion of the target bit is indicated by the `crossed-circle'.  The CNOT gate can be extended to include two or more input states, an example of which is the CCNOT or Toffoli gate, shown in figure \ref{fig:toffoli}.  Similarly the Toffoli gate inverts the target bit if both control bits are equal to $1$.  Other gates of interest are the inversion or NOT gate (figure \ref{fig:invertswap}(a)) and the swap gate (figure \ref{fig:invertswap}(b)) which just swaps the input states $|a\rangle$ and $|b\rangle$.  If the dependance on the control bit of any of these gates is reversed, this is indicated by a open circle as shown in figure \ref{fig:invertedcnot}.  The fact that the control bits for both the CNOT and Toffoli gates have not changed after the operation is a result of the reversibility of these gates.

While these reversible gates are essentially a classical concept, the power of quantum computing arises when they are applied to quantum superposition states where a qubit can be in some superposition of $0$ and $1$.  In order to prepare these superposition states, the quantum computer must be able to perform so called `rotations' to set up an arbitrary superposition.  The result of this is that the control and target qubits for the quantum version of these gates can be in some unknown state which is only determined by measurement of the qubit.  The measurement process returns one of two possibilities where the probability of returning a given state is the square of the amplitude of that state in the superposition.  The application of this kind of measurement is indicated by the symbol in figure \ref{fig:invertswap}(c) and is termed a destructive measurement as all information about the quantum superposition is lost.

For each circuit in this paper, $|S\rangle_{i}$ is designated the on-site Ising spin ($\mathbf{s}_{i}$) where $S_{i}=\frac{\mathbf{s}_{i}+1}{2}$ and $|S'\rangle_{i}$ is the new on-site spin after application of the quantum circuit.  This allows a simple correspondence between the Ising states and the qubit encoding, giving `spin-up' ($+1$) as binary state $1$ and `spin-down' ($-1$) as state $0$.

\section{1D Ising model}\label{1dising}
\begin{figure} [htb]
\centering{\includegraphics[width=6cm]{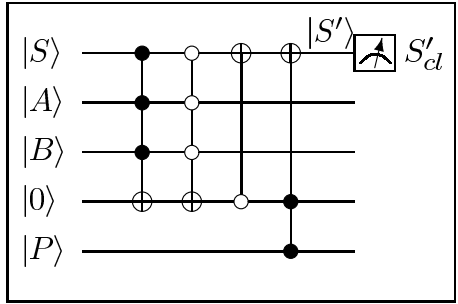}} \caption
{Quantum circuit for the evolution of the on-site spin $|S\rangle$ for the 1D Ising model\label{fig:1Dcct}}
\end{figure}

For a particular on-site spins $S_{i}$ of the 1D Ising model, the neighbouring spins are designated $A$ and $B$.  As the spins only take discrete values the change in energy due to a single spin flip can only take a finite set of values ($\Delta E=0J$, $\pm4J$), as long as there is no global field.  This means the Metropolis method can be implemented using the circuit given in figure \ref{fig:1Dcct}, where $|P\rangle$ is a qubit initialised in the state $|P\rangle=\sqrt{P}|1\rangle+\sqrt{1-P}|0\rangle$ with $P=e^{-\frac{4J}{T}}$ giving the appropriate probability
for the given temperature and coupling strength.  The circuit also has an `ancilla' qubit which is initialised with the state $|0\rangle$ and is used for temporary storage during the computation.

The general algorithm steps are as follows
\begin{enumerate}
    \item   Initialise on-site spin for each node ($|S\rangle_{i}$) based on the initial state of the Ising lattice
    \item   Initialise input states $A$ and $B$ based on the value of the neighbouring spins using the streaming rule (see section \ref{streaming})
    \item   Apply quantum circuit gate sequence to the entire computer simultaneously
    \item   Measure resultant spin for each node
    \item   Use resultant spin to re-initialise neighbouring nodes
    \item   Repeat until equilibrium is reached
\end{enumerate}

The parameters of interest (mean magnetisation for instance) can be calculated once the output of each node is measured.  In figure \ref{fig:1Dcct} the circuit requires 5 qubits and 4 multi-qubit gates though it is possible to reduce the number of qubits by rewriting the circuit in terms of two-qubit gates and single qubit rotations\cite{Barenco:95}.  The choice of which two-qubit gates to use for this decomposition is largely a matter of which gates are most suitable for a given implementation. In this form the circuit will require at least $2d+1$ qubits for a $d$-dimensional lattice to represent the on-site spin and its nearest neighbours.

The truth-table given in figure \ref{fig:1DTT} shows that this circuit is equivalent, once measured, to the classical Metropolis method. This table gives the output of the quantum circuit $|S'\rangle$ for each input state $ASB$ as well as the classical result $S'_{cl}$ and its probability of acceptance $p_{cl}$.  The result of the circuit given in \ref{fig:1Dcct}, once the state $|S'\rangle$ is measured, is identical to the classical result.  All but two of the entries in this table are exactly equivalent to the classical metropolis result, where a spin flip is accepted as long as the change in energy is zero or negative.  The first (last) entry differs in that the result is a superposition state with a probability $P$ of obtaining a spin-up (spin-down) result and a corresponding probability $1-P$ of obtaining the opposite result.  This is equivalent to the random number generation step of the metropolis algorithm where a configuration is rejected or accepted based a weighted probability.

\begin{figure}[htb]
\begin{center}
    $\begin{array}{|c||c||c|c|}
        \hline ASB &|S'\rangle &S'_{cl} &p_{cl} \\
        \hline \downarrow\downarrow\downarrow &\sqrt{P}|\uparrow\rangle+\sqrt{1-P}|\downarrow\rangle    &\downarrow &1-P \\
        \cline{3-4} &    &\uparrow &P \\
        \hline \downarrow\downarrow\uparrow &|\uparrow\rangle    &\uparrow &1 \\
        \hline \downarrow\uparrow\downarrow &|\downarrow\rangle    &\downarrow &1 \\
        \hline \downarrow\uparrow\uparrow &|\downarrow\rangle    &\downarrow &1 \\
        \hline \uparrow\downarrow\downarrow &|\uparrow\rangle    &\uparrow &1 \\
        \hline \uparrow\downarrow\uparrow &|\uparrow\rangle    &\uparrow &1 \\
        \hline \uparrow\uparrow\downarrow &|\downarrow\rangle    &\downarrow &1 \\
        \hline \uparrow\uparrow\uparrow &\sqrt{P}|\downarrow\rangle+\sqrt{1-P}|\uparrow\rangle    &\uparrow &1-P \\
        \cline{3-4} &    &\downarrow &P \\
        \hline
    \end{array}$
    \\Note: $P=e^{-\frac{4J}{T}}$
\end{center}

\caption{Truth-table for the quantum circuit in \ref{fig:1Dcct}, showing the equivalence to the classical Metropolis algorithm once the superposition state $|S'\rangle$ is measured}\label{fig:1DTT}
\end{figure}

\section{2D Ising model}\label{2dising}
\begin{figure} [htb]
\centering{\includegraphics[width=13cm]{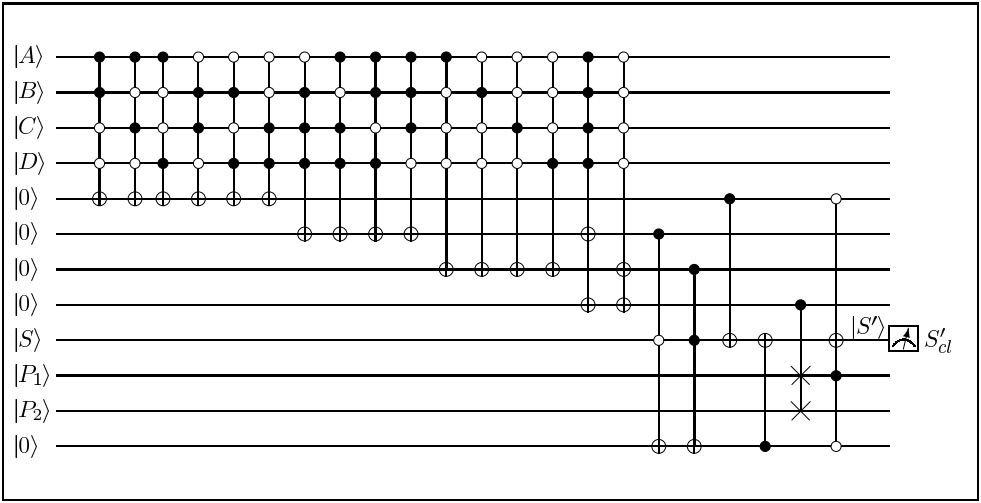}} \caption
{Quantum circuit for 2D Ising model\label{fig:2Dcct}}
\end{figure}

Using the same approach, the Metropolis method for the 2D Ising model can be implemented using the circuit given in figure \ref{fig:2Dcct}.

Inputs A-D are the neighbouring spins and $P_{1}$ and $P_{2}$ are the probabilities $P_{1}=e^{-\frac{4J}{T}}$ and
$P_{2}=e^{-\frac{8J}{T}}$ respectively.  These correspond to qubits initialised in the states $|P_{1}\rangle=\sqrt{P_{1}}|1\rangle+\sqrt{1-P_{1}}|0\rangle$ and $|P_{2}\rangle=\sqrt{P_{2}}|1\rangle+\sqrt{1-P_{2}}|0\rangle$.

The circuit given for the 2D Ising model is not optimal and can in fact be realised with considerably fewer and simpler gates using a `full-adder' type configuration.  It is presented in this form as it is more apparent which sections of the circuit operate for $\Delta E=0J$, $\pm4J$ and $\pm8J$, making the operation of the circuit more obvious.  If this circuit were to be realised experimentally, it could be broken down into the appropriate two qubit operations which suited the particular implementation~\cite{Barenco:95}.  This can result in considerable simplification, depending on the details of the experimental system.

\section{Streaming and parallelisation}\label{streaming}
\begin{figure} [htb]
\centering{\includegraphics[width=10cm]{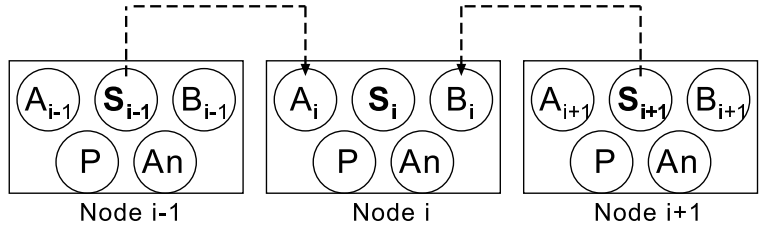}} \caption
{Each node is initialised with the on-site spin $S_{i}$ and copies of the neighbouring on-site spins $S_{i-1}$ and $S_{i+1}$\label{fig:streaming}}
\end{figure}

The streaming process for the 1D Ising chain is illustrated in figure \ref{fig:streaming}.  Each node is initialised using the value for the on-site spin $S_{i}$ and copies of its nearest neighbours $A_{i}=S_{i-1}$ and $B_{i}=S_{i+1}$.  The superposition qubit ($P$) and the ancilla qubit ($An$) for each node are re-initialised every iteration with the same value, as discussed in section \ref{1dising} and \ref{2dising}.

\begin{figure} [htb]
\centering{\includegraphics[width=4cm]{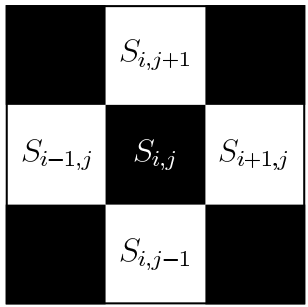}} \caption
{The checkerboard update scheme with the spin value $S_{i,j}$ and its four nearest neighbours $A=S_{i,j+1}$, $B=S_{i-1,j}$, $C=S_{i,j-1}$ and $D=S_{i+1,j}$ illustrated\label{fig:checker}}
\end{figure}

The streaming rule for the 2D Ising lattice is similar to the 1D case.  As discussed earlier, all nodes cannot be updated simultaneously otherwise the lattice will oscillate and not reach a stable state.  A checkerboard update scheme can be used instead, as illustrated in figure \ref{fig:checker}.  Each `black' node is initialised with its spin value $S_{i,j}$ and its four nearest neighbours $A=S_{i,j+1}$, $B=S_{i-1,j}$, $C=S_{i,j-1}$ and $D=S_{i+1,j}$.  The new value of $S$ is calculated and then the same process is followed for the `white' nodes.  Once this process is completed, every node in the lattice has been updated.

As a result of the reversibility of quantum circuits and the fact that the target qubits are left unchanged, the values of the neighbouring spins $A$-$D$ are preserved after the computation.  This means that no additional storage space is required to store the spins that are not being modified at each evolution (assuming error free operation).  It also means that the value of an arbitrary neighbour (ie: $A$) can be used to initialise the spin $S$ at the next time step.  This results in only half as many nodes being required to simulate the lattice.  This algorithm therefore requires $\frac{N}{2}$ computational nodes for $N$ Ising spins. In this situation two evolutions of the quantum circuit are required to update every node in the lattice once.  An alternative solution is to include a parity bit which controls whether the rest of the circuit functions or not.  The lattice is then initialised so that the parity bits alternate across the lattice and they are inverted at each evolution. 

\section{Possible implementations}\label{implement}
The algorithm, as discussed, is independent of implementation and would therefore be suited to a range of different quantum computing architectures.  Of particular interest is that the final state of the spin after each evolution collapses to either a $1$ or a $0$ when measured.  Existing diffusion based type-II algorithms require the probability amplitudes to be measured at each step, which requires either time or ensemble averaging.  This cellular automaton based algorithm, on the other hand, does not require the measurement of the full ensemble of states and is therefore suited to solid state or ion-trap architectures.

One important consideration is that the accuracy with which the temperature of the system can be controlled is directly related to the accuracy with which the superposition state $|P\rangle$ can be created.  The amplitude ($p=\sqrt{P}$) of this superposition is given by $p=e^{-2/T}$ for the 2D case and $p=e^{-1/T}$ for 1D. The accuracy with which this superposition can be created and controlled puts a lower bound on the temperature resolution the system can simulate. This is especially important for architectures where only a binary output is measured, as the superposition must be created using single qubit rotations.  The accuracy in the temperature is therefore related to the accuracy of the rotation gate.  By using standard uncertainty analysis, the maximum allowable error rate ($\delta p$) can be calculated for a given temperature resolution using equation \ref{equ:pError}.

\begin{equation} \label{equ:pError}
\delta p=\left|\frac{dp}{dT}\right|\delta T
\end{equation} 

\begin{figure} [htb]
\centering{\includegraphics[width=12cm]{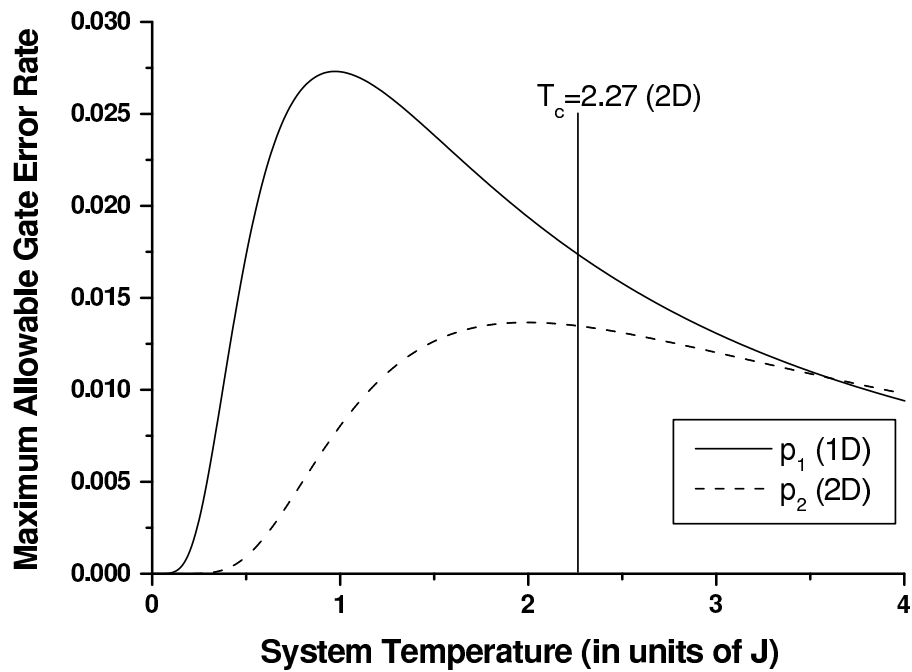}} \caption
{Maximum allowable error rate required for a temperature resolution of $\delta T=0.1$, as a function of Ising lattice temperature\label{fig:gateacc}}
\end{figure}

Figure \ref{fig:gateacc} shows the maximum allowed gate error rate for a single qubit rotation ($\delta p$) needed to achieve a temperature resolution ($\delta T$) of 0.1, as a function of system temperature ($T$).  The curves $p_{1}$ and $p_{2}$ correspond to the maximum allowed gate error rate ($\delta p$) for the 1D and 2D systems respectively.  For $T<0.5$ the required gate accuracy is effectively infinite but in this region the simple Ising system displays trivial behaviour.  For the region of interest ($1.5<T<4$), a maximum error rate of $\delta p=0.01$ is sufficient for 1D and 2D calculations with a temperature resolution of 0.1.  The two-qubit gates must also be controlled with a comparable resolution, though the net effect of these gates should be at most linear in the number of gates.

\section{Ensemble streaming}\label{ensemble}
\begin{figure} [htb]
\centering{\includegraphics[width=12cm]{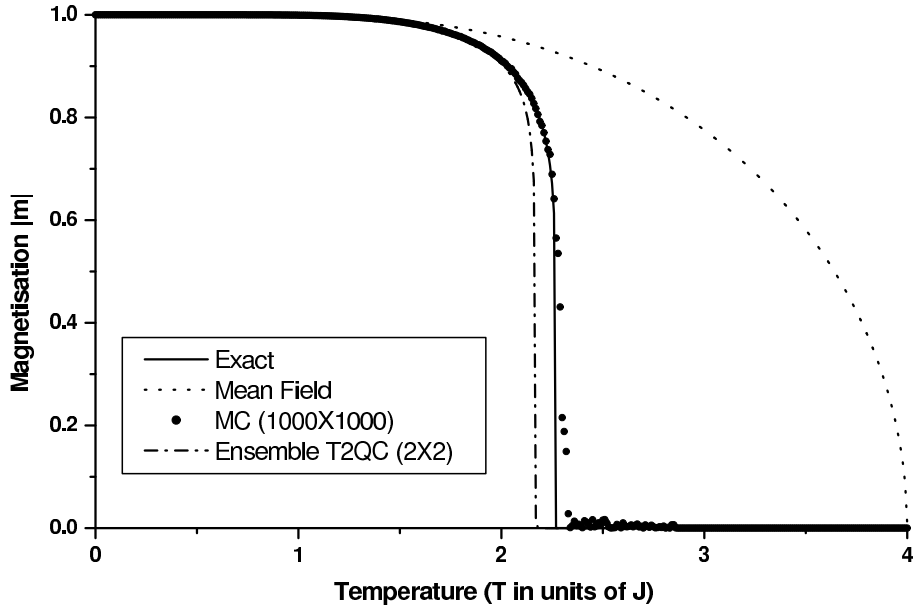}} \caption
{Magnetisation as a function of temperature for the 2D Ising model.  The `Exact' solution is the analytic solution for any infinite lattice, `Mean field' is the mean field approximation for an infinite lattice, `MC' is a classical Metropolis monte-carlo run using a $1000 \times 1000$ lattice and `Ensemble T2QC' is a simulated run of the Ising algorithm for a T2QC using ensemble streaming on 4 nodes\label{fig:solcomp}}
\end{figure}

The algorithm detailed so far is formally equivalent to the classical metropolis algorithm for the Ising model.  An interesting side note involves determining what would happen if the measurement step were to be replaced with an ensemble measurement.  An ensemble measurement consists of measuring many copies of the system and determining the magnitude of the probability distribution for each qubit.  This distribution can then be used to re-initialise the other spins, ready for the next iteration of the algorithm.  This situation corresponds to either time averaging over many runs or ensemble averaging, as is the case with NMR based implementations, and will be referred to as `ensemble streaming'.  This is especially interesting as the only type-II quantum computer in existence at the time of writing is based on the NMR architecture and therefore uses ensemble streaming\cite{Pravia:02}\cite{Pravia:03}.  The result of this process is that the spins no longer take on definite integer values but can have continuous values depending on their neighbours and the temperature dependant function.  The magnetisation as a function of temperature is shown for this ensemble streaming (Ensemble T2QC) for the 2D Ising model (the dot-dash line in figure \ref{fig:solcomp}).

By allowing the full ensemble of states to propagate through the lattice, the model now bears some similarity to a mean-field style approximation.  As each node can vary continuously from 0 to 1, the deviation from the mean spin value quickly decays to zero for any one spin site.  After a few iterations, each node on the lattice takes on the same expectation value, resulting in a smooth curve even for a very small lattice of four spins.  As a result, the curve for any number of spins is found to be the same as for the four spin case.  In addition the ensemble version converges much faster than the classical Metropolis algorithm (less than 5 iterations per temperature point).  The curve (figure \ref{fig:solcomp}) corresponds to the system being heated from the ground state up to a state where every configuration is equally likely, which is equivalent to the randomised state.  If the system was then cooled again, it tends to stay completely randomised and therefore the zero magnetisation state is the stable solution all the way back to zero temperature.  This is a consequence of allowing the spins to be continuous (an ensemble of states) rather than taking discrete values and is a non-physical solution. 

For comparison, the solution for the ensemble T2QC is plotted in figure \ref{fig:solcomp} with the analytic solution due to Onsager~\cite{Onsager:44} for an infinite lattice, a classical Metropolis MC run using a $1000\times1000$ point lattice and the classical mean field solution for an infinite lattice~\cite{Chandler:87}.  The classical Metropolis MC run was performed using a temperature step size of 0.01 with a minimum of 20 and a maximum of 10000 iterations per temperature point.  This includes $1000^2$ spin flips per iteration with each spin flip requiring the generation of a random number, the calculation of the change in energy and the evaluation of an exponential weighting function.  By comparison, for the standard T2QC algorithm with one-shot measurement all these steps are performed through the evolution of the circuit in figure \ref{fig:2Dcct}.  This occurs for each spin in the lattice simultaneously which means that the circuit need only be applied once per iteration.  The number of iterations required for convergence using the one-shot measurement algorithm is similar to the classical algorithm.

The ensemble measurement algorithm underestimates the critical temperature, as can be seen in figure \ref{fig:solcomp}, giving $T_{c}\approx2.109$, compared to the analytic solution of $T_{c}=2.269$~\cite{Onsager:44}.  The form of the ensemble solution is a polynomial approximation rather than the accepted power law behaviour.  This is due to the fact that as the true Ising lattice approaches the critical temperature, large clusters of spins form in random locations in the lattice, resulting in non-zero magnetisation.  The ensemble measurement algorithm, on the other hand, effectively averages over all possible configurations.  As the clusters occur in random locations their effects cancel out, resulting in a zero average magnetisation.  This discrepancy accounts for the difference in critical temperature and gives some insight into the regime where clustering is an important effect.  It is expected that a large cluster generalisation of this algorithm would improve the critical behaviour near $T_{c}$ as the forming of clusters would be more accurately modelled.

\section{Conclusion}\label{conc}
A novel algorithm for a type-II quantum computer has been presented for simulating the Metropolis Monte-Carlo method for the Ising model in one and two dimensions.  The algorithm requires only $N/2$ nodes to simulate $N$ spins and is formally equivalent to a probabilistic cellular automaton formulation of the Metropolis method for the Ising model.  It requires a maximum of 5 qubits per node for the 1D case and 12 qubits per node for the 2D case, though this is an upper bound for arbitrary interactions.  This compares well with other classical implementations which require many more classical bits for random number generation.  It also demonstrates that other cellular automata models may be simply reformulated to run on a type-II quantum computer.  This extends the range of applications for a type-II quantum computer beyond those based on the lattice Boltzmann hydrodynamic equations, such as diffusion.  Additionally, the effect of running this algorithm on a type-II quantum computer of the type demonstrated experimentally, using ensemble streaming, was studied.  While this ensemble streaming variation of the algorithm was found to replicate the phase change of a two dimensional Ising lattice, it is unable to accurately describe the effects of clustering near this transition.

\section{Acknowledgements}\label{Ack}
This work was supported by the Australian Research Council.  The authors wish to acknowledge helpful discussions with S.~P.~Russo and C.~J.~Wellard.



\bibliography{t2ising}

\end{document}